\documentclass[twocolumn,showpacs,amsmath,amssymb,10pt]{revtex4}
\usepackage{graphicx}
\usepackage{dcolumn}
\usepackage{bm}

\def\tr{\mbox{tr}}

\def\bea{\begin{eqnarray}}
\def\eea{\end{eqnarray}}

\begin{document}

\title{Minimum-error state discrimination constrained by the no-signaling principle}

\author{Won-Young Hwang}
\affiliation{Department of Physics Education, Chonnam National University, Gwangju 500-757, Korea \\}
\author{Joonwoo Bae}
\email{bae.joonwoo@gmail.com}
\affiliation{School of Computational Sciences, Korea Institute for Advanced Study, Seoul 130-012, Korea}

\date{\today}


\begin{abstract}
We provide a bound on the minimum error when discriminating among quantum states, using the no-signaling principle. The bound is general in that it depends on neither dimensions nor specific structures of given quantum states to be discriminated among. We show that the bound is tight for the minimum-error state discrimination between symmetric (both pure and mixed) qubit states. Moreover, the bound can be applied to a set of quantum states for which the minimum-error state discrimination is not known yet. Finally, our results strengthen the quantitative connection between two no-go theorems, the no-signaling principle and the no perfect state estimation.
\end{abstract}


\maketitle

\section{Introduction}

The no-signaling principle, one of the no-go theorems in quantum theory, asserts that information cannot be sent faster than light. In fact, the fundamental theorem can be utilized to derive some limitations on performance of quantum operations. One possible way to make use of the no-signaling condition is to incorporate a quantum operation to a communication scenario, in which the operation should be limited such that any information cannot be sent faster than light.

It was shown that quantum cloning can be incorporated with a communication scenario. By the no-signaling principle, an upper bound to the optimal cloning fidelity was obtained in the $1\rightarrow 2$ universal quantum cloning, and it turns out that the bound coincides with the optimal cloning fidelity \cite{nocloning-gisin}.  It was also shown that quantum state discrimination can be incorporated with a communication scenario, to which the no-signaling principle could be applied. Upper bounds to the optimal ones were obtained in cases of unambiguous discrimination \cite{steph1}, minimum-error (two-states) discrimination \cite{hwa,bae}, and maximum confidence state discrimination \cite{steph2}. It was shown that the upper bounds are tight, so the optimal bounds are immediately implied by the no-signaling principle. Note that postulates of quantum theory, such as positivity of quantum operations and measurement postulates, are assumed throughout.

The present article contributes to a general and novel connection between the no-signaling principle and the minimum-error state discrimination among multiple quantum states. From the no-signaling constraint, we derive an explicit formula that non-signaling probability distributions have to fulfill, by which a bound to the minimum-error state discrimination among quantum states can be obtained. We show that the bound coincides with the optimal ones for some cases where the minimum-error state discrimination is known, such as pure or mixed symmetric qubit states. Note that the bound can be compared with other known bounds, for instance in \cite{monta,tyson1,qiu, qiu0}.

The importance of this work is twofold. On one hand, in the side of applications, our work provides a systematic way to compute an upper bound to the success probability in the minimum-error state discrimination \cite{helstrom,geo,herzog,jar,y,h1,h2}. Although quantum state discrimination is one of basic operations to estimate and characterize capabilities of tasks in Quantum Information Theory, the optimal minimum-error discrimination is known for few cases such as two-state or symmetric states discrimination \cite{helstrom, geo, herzog, jar}. Here, our method gives a bound to the minimum-error state discrimination among high-dimensional states for which the optimal discrimination is not known yet.

On the other hand, in the fundamental point of view, our work strengthens the quantitative connection between two no-go theorems, the no-signaling and no perfect state estimation. First, from two facts that i) the no perfect cloning holds for all non-signaling theories having Bell violations \cite{acin} and ii) quantum theory is non-local \cite{bell}, it is clear that the no-signaling principle in quantum theory immediately implies that perfect quantum cloning is not possible \cite{no cloning}. To be quantitative, it was shown that the optimal cloning fidelity can be obtained by the no-signaling principle \cite{nocloning-gisin}. Moreover, quantum cloning generally converges to the optimal state estimation in the asymptotic limit \cite{jbae}. However, the connection between the state estimation and the no-signaling is only qualitative so far through the above mentioned connection: the no-signaling principle implies that the perfect cloning is not possible, and therefore the state estimation cannot be perfectly done either. Along these lines, the quantitative connection has been shown in the minimum-error state discrimination between two pure states \cite{hwa} and, generally, two mixed ones \cite{bae}.

In this work, we provide a general upper (lower) bound to success (error) probability in discriminating among quantum states by the no-signaling principle. In Sec.\ref{main}, we derive the bound using the no-signaling constraint. In Sec.\ref{qubit}, the bound is shown to be tight for cases where the minimum-error state discrimination is known. Finally, an example of discriminating among high-dimensional states is provided in Sec.\ref{qudit}.

\section{Main result}
\label{main}

Let us first introduce the communication scenario to which the minimum-error state discrimination can be incorporated. Two parties, Alice and Bob, are separated by a distance such that local actions cannot affect the other side. Suppose that they share copies of entangled states
\bea |\psi\rangle_{AB} = \sum_{n}\sqrt{\lambda_{n}} |u_{n}\rangle_{A} |v_{n}\rangle_{B}, \label{ab}\eea
where $|u_{n}\rangle_{A} \in \mathcal{H}_{A} $ and $|v_{n}\rangle_{B}\in \mathcal{H}_{B}$, finite dimensional Hilbert spaces, $\mathcal{H}_{A}$ and $\mathcal{H}_{B}$. Now, Alice applies general measurement, Positive-Operator-Valued-Measure (POVM) $M_k = \{ M_k^{j} \} $, a complete measurement, i.e. $\sum_{j} M_k^{j} = I$. Here, let the choice of POVM, denoted by the value $k$, be the message that Alice wishes to send to Bob. The measurement by Alice, say $M_k$, prepares the decomposition $\rho_{B,k}$ on Bob's side,

\bea \rho_{B,k} = \sum_{j} p_{k}^{j} ~\rho_{B,k}^{j}, \label{bobs} \eea
where $p_{k}^{j} = \tr[(M_{k}^{j}\otimes I_{B}) |\psi\rangle_{AB}\langle\psi| ]$ and $\rho_{B,k}^{j} = (p_{k}^{j})^{-1}\tr_{A}[(M_{k}^{j}\otimes I_{B}) |\psi\rangle_{AB}\langle\psi|]$. Since Alice's measurement $M_{k}$ is complete for each $k$, Bob's ensembles are all identical,
\bea \rho_{B,k} = \rho_{B,l}, ~~~\forall k \neq l, \label{con} \eea
while they are in general different decompositions. Moreover, Alice can generate any decomposition as long as the ensemble is identical, as it is known by Gisin-Hughston-Jozsa-Wootters theorem \cite{Gis89,Hug93}.

It is clear that, since the ensembles are identical, Bob never knows which measurement has been applied by Alice. The no-signaling principle would be then violated if Bob could discriminate among the different decompositions of his ensemble without any further communication with Alice. In what follows, using the no-signaling constraint, we derive the condition that has to be fulfilled in the state discrimination.

\subsection{Constraint by no-signaling principle}

Suppose that Bob applies a detector to make a guess of which measurement has been applied by Alice. Note that we do not know all properties of the detector but the probability distributions. In fact, the detector can be thought of being in a black box from which one can only know the input-output list. The detector of Bob works as follows. Once Alice applies measurement $M_{l}$ on her system to send a message $l$, Bob's detector gives an outcome $k$ with some probability $D(k| A_{l})$, from which he makes a guess that $M_{k}$ has been applied. Since they are probabilities, for each $l$ the normalization condition holds
\bea \sum_{k} D(k|A_{l}) =1.\label{unit}\eea
Then, the probability that the detector tells the correct result is, $\sum_{k} p(A_{k}) D(k|A_{k})$, where $p(A_{k})$ is the probability that Alice applies measurement $M_{k}$. Throughout the paper, we restrict to the case where $p(A_{k})$ is the same for all $k$, i.e. a measurement $M_{k}$ for some $k$ is applied randomly. If there are $N$ possible measurements by Alice, $p(A_{k}) = 1/N$ for all $k$.

The no-signaling principle can be imposed as follows. Suppose that the detector works too well such that the success probability is better than random,
 \bea \sum_{k} \frac{1}{N} D(k|A_{k}) > \frac{1}{N}.\label{ns} \eea
Together with Eqs. (\ref{unit}) and (\ref{ns}), we get
\bea \sum_{k}[D(k|A_{k}) - D(k|A_{l})] > 0. \label{new} \eea
Clearly, it holds for at least a single value $k$ that
\begin{equation}
D(k|A_{k})- D(k|A_{l})>0.
\label{A}
\end{equation}
That is, there exists a value $k$, Alice's encoding $A_{k}$, to which Bob's detector responds better than other ones $A_{l}(\neq A_{k})$. This immediately implies that a faster-than-light communication is possible in the following way. If Alice wants to send a message $0$ ($1$), she repeatedly performs her measurements such that the $k$-th ($l$-th) decomposition is generated at Bob's site. Bob repeatedly performs his measurements. Then Bob can discriminate between the two cases by observing how frequently his detector gives the $k$-th outcome, by Eq. (\ref{A}). Thus Bob can decode a one bit message. We then have the following proposition.  \\

\textbf{\emph{Proposition 1.}} \emph{Suppose that two parties Alice and Bob sharing copies of entangled states in Eq. (\ref{ab}), Alice prepares different ensemble decompositions on Bob's side by applying complete measurement $M_{l}$ to her systems with the equal probability $1/N$, where $l = 0,\cdots,N-1$. Let $D(k|A_{l})$ be the probability that a detector of Bob answers the $k$-th decomposition when Alice has applied measurement $M_{l}$.Then, the no-signaling principle constrains the success probability,
\bea \sum_{k} D(k| A_{k}) \leq 1. \label{nos}\eea
Note that neither dimensions nor particular structures of given states are assumed.}

\subsection{The no-signaling constraint is imposed on the state-discrimination}

Now we apply the condition in Eq. (\ref{nos}) obtained by the no-signaling constraint to minimum-error discrimination among states in $\{ \rho_{k} ~\|~ k=0, \cdots, N-1 \}$. The idea is to construct the identical ensemble having different decompositions $\{ \rho_{B,k}~ \|~ k=0,\cdots, N-1 \}$ such that each decomposition $\rho_{B,k}$ contains $\rho_{k}$. The condition in Eq. (\ref{nos}) then gives a bound to the minimum-error discrimination among states in $\{ \rho_{k} ~\|~ k=0, \cdots, N-1 \}$.

Suppose that, as it is shown in Eq. (\ref{bobs}) by Alice's measurement, each state $\rho_{k}$ is contained in the ensemble decomposition $\rho_{B,k}$ with probability $p_{k}$
\bea \rho_{B,k} = p_{k} \rho_{k} + \sum_{k'\neq k} q_{k'} \sigma_{k'}. \label{decom}\eea
Note that in the above we are not interested in the state $\sum_{k'\neq k} q_{k'} \sigma_{k'}$, which only helps to construct the identical ensemble. Because the ensembles corresponding to $\rho_{B,k}$ are identical (see Eq. (\ref{con})), we have
\bea p_{k} \rho_{k}  + \sum_{k'\neq k} q_{k'} \sigma_{k'} ~~= ~~p_{l} \rho_{l}  + \sum_{l'\neq l} q_{l'} \sigma_{l'}. \label{con2} \eea

Let us now consider Bob's detector designed to discriminate among states in $\{ \rho_{k} ~\| ~ k= 0,\cdots,N-1 \}$. The detector can also be thought of being in a black box from which Bob only knows the input-output list. Let $P(k|\rho_{l})$ denote the probability that Bob's detector gives an outcome $k$ (meaning that Bob guesses a state $\rho_{k}$ is given) when $\rho_{l}$ is actually given \cite{note}. It follows from Eq. (\ref{decom}) that
\bea p_{k} P(k|\rho_{k}) \leq D(k|A_{k}), \label{dp} \eea
since Bob's detection consists of the contributions both by the state $\rho_{k}$ with probability $p_{k}$ and by other states in the ensemble such as  $\sum_{k'} q_{k'} \sigma_{k'}$. From Eq. (\ref{dp}) and  Eq. (\ref{nos}) in the proposition, the no-signaling condition implies that $ \sum_{k} p_{k} P(k|\rho_{k}) \leq 1$. We can then summarize in the following corollary. \\

\emph{\textbf{Corollary 1.} Suppose that a set of states $\{\rho_{k}\}$ for $k=0,\cdots, N-1$ is given with a priori probabilities $1/N$, to be discriminated between. If one can construct a set of identical ensembles $\{\rho_{B,k}\}$ where each ensemble $\rho_{B,k}$ is decomposed such that $\rho_{k}$ is contained in the ensemble with probability $p_{k}$ (see Eqs. (\ref{con}) and (\ref{decom})), then it must be fulfilled by the no-signaling constraint that
\bea \sum_{k} p_{k} P(k|\rho_{k}) \leq 1. \label{bd} \eea
Here $P(k|\rho_{k})$ is the probability that Bob's detector answers the $k$-th state $\rho_{k}$ when $\rho_{k}$ is prepared. }\\

Note that the corollary can be directly applied to the success probability in the state discrimination.


\subsection{Application to a set of quantum states}

Let us now explain how to apply the corollary to the minimum-error state discrimination among multiple quantum states, $\{\rho_{0}, \rho_{1},\cdots, \rho_{N-1}\}$. First, one has to construct an identical ensemble in different decompositions, $\{\rho_{B,0},\rho_{B,1},\cdots,\rho_{B,N-1}\}$ (i.e. $\rho_{B,k} = \rho_{B,l}$ for all $k\neq l$), such that each decomposition contains one of the states to be discriminated among,
\bea \rho_{B,0} & = & p_{0}  \rho_{0} + \cdots \nonumber \\
\rho_{B,1} & = & p_{1}  \rho_{1} + \cdots \nonumber \\
 & \cdots &  \nonumber \\
\rho_{B,N-1} & = & p_{N-1}  \rho_{N-1} + \cdots.  \label{ensem} \eea
It is then straightforward to apply the corollary to obtain a bound to state discrimination. The goal here is to minimize the error, or equivalently to maximize the success probability. The success probability is, $(1/N) \sum_{k} P(k|\rho_{k})$, on average since we are considering the case where each state to be discriminated is generated with equal probabilities, $1/N$. Now let us assume that different decompositions, denoted by $C$, of an identical ensemble are given. The success probability is then constrained from the corollary. Therefore, the bound to the success probability $P_{s}^{C}$ for a given decomposition $C$ is,

\bea
P_{s}^{C} =  \max_{\sum_{k} p_{k} P(k|\rho_{k}) \leq 1} \big\{ ~\frac{1}{N} \sum_{k} P(k|\rho_{k})\big\}.\label{pd}
\eea
Note that the no-signaling principle must be fulfilled for any ensemble decompositions under consideration. Hence the success probability $P_{s}$ in the state discrimination is upper bounded as follows,
\bea P_{s} \leq  \min_{C} P_{s}^{C}, \label{res} \eea
where the minimization is taken over all decompositions that can be constructed in Eq. (\ref{ensem}). Note also that, given a decomposition $C$, the success probability $P_{s}^{C}$ in Eq. (\ref{pd}) gives a general upper bound to the success probability in the minimum-error state discrimination. For the particular case of the ensemble decomposition where $p_{k}:=p $ for all $k$, the bound to state discrimination is simplified, i.e. the success probability is upper bounded, \bea P_{s} \leq (Np)^{-1} \label{form}.\eea In this case, although the decomposition may not be the optimal one, no maximization step is required. Therefore, constructing the identical ensembles immediately provides an upper bound to the success probability.

Note that the discrimination bound in Eq. (\ref{res}) assumes neither dimensions of given states nor any structure among states $\{ \rho_{k}\}$ to be discriminated. It only assumes that an identical ensemble can be in different decompositions $\rho_{B,k}$ (see Eq. (\ref{con2})). The upper bound to the success probability can be compared with known bounds shown in Refs. \cite{monta,qiu,tyson1}. In the next section, our bound is shown to be tight for $N$ symmetric qubit states for which the minimum-error state discrimination is known \cite{geo}. We will explicitly compute the upper bound to success probability from Eq. (\ref{form}), and compare with the optimal bound.

\section{Application to $N$ symmetric qubit states}
\label{qubit}

We consider minimum-error discrimination among $N$ symmetric qubit (pure and mixed) states for which the optimal measurement is known. We first apply the bound in Eq. (\ref{res}) to $N$ symmetric qubit states. Here, the symmetric states $\rho_{j}$ are characterized by a unitary operator $U$ that satisfies, $U^{N} = I$, and \bea U\rho_{j}U^{\dagger} = \rho_{j+1}, ~~\forall j=0, 1 , \cdots , N-1. \label{sym}\eea They are also called as geometrically uniform states \cite{geo}. For qubit states, the unitary operation $U$ is a rotation operator with respect to an axis in the Bloch sphere. It will be shown that, remarkably, bounds obtained by the no-signaling principle turn out to be tight.

\begin{figure}
\begin{center}
  \includegraphics[width= 6.7 cm]{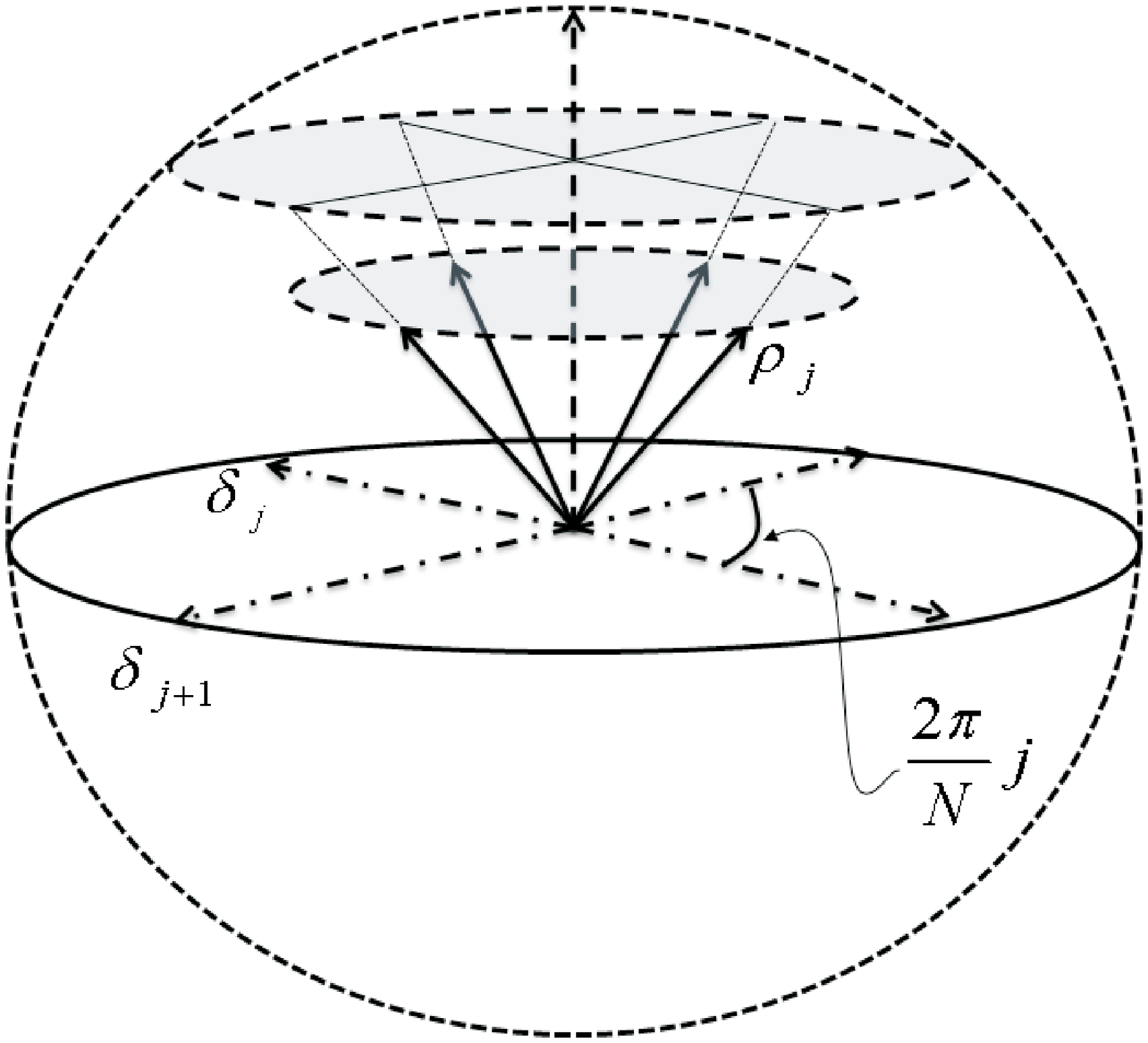}
  \caption{Both $\rho_{j}$ and $\delta_{j}$ form $N$ symmetric states with the unitary operation $U$, the rotation by $2\pi j/N$ with respect to the $\hat{z}$-axis. The ensemble $\rho_{B,j}$ in Eq. (\ref{ben}) points to the $\hat{z}$-axis in the Bloch sphere, having different decompositions, see Eq. (\ref{ben}) . }\label{fig1}
\end{center}
\end{figure}

\subsection{Bound from the no-signaling principle}

To apply the corollary to state discrimination, one has to first construct identical ensembles. Let us first start with the fact that a qubit state can be expressed as, \bea \rho(\vec{n}) = (I +\vec{n} \cdot \sigma)/2, \label{state} \eea where $\vec{n} = |\vec{n}| (\sin\theta\cos\varphi,\sin\theta\sin\varphi,\cos\theta)$, and $|\vec{n}| = 1$ for pure states, see Fig (\ref{fig1}). $N$ symmetric qubit states that we are to discriminate among can be parameterized as follows: \bea\rho_{j} & = & (I +\vec{n}(\rho_{j}) \cdot \vec{\sigma})/2, ~~~ j =0,\cdots,N-1, \nonumber \\
\vec{n}(\rho_{j}) & = & |\vec{n}| (\sin\theta \cos\frac{2\pi}{N}j, \sin\theta \sin\frac{2\pi}{N}j, \cos\theta), \label{nqstate} \eea with the unitary operation, denoted by $V$, that rotates a state $\rho_{j}$ to $\rho_{j+1}$ by $2\pi/N$ with respect to the $z$-axis. For each state $\rho_{j}$ in (\ref{nqstate}), we introduce the state $\delta_{j}$ lying on the half plane of the Bloch sphere, \bea \delta_{j} & = & (I + \hat{n}(\delta_{j}) \cdot \vec{\sigma})/2, ~~j=0,\cdots,N-1 \label{delta} \\  \hat{n}(\delta_{j}) & = & (-sgn( \sin \theta )\cos\frac{2\pi}{N}j, - sgn( \sin \theta )\sin\frac{2\pi}{N}j,0). \nonumber \eea Then, the identical ensembles can then be constructed such that their Bloch vectors point to the same point in the $z$-axis of the Bloch sphere:  \bea \rho_{B,i} & = & \rho_{B,j}, ~~~~\forall ~i\neq j\nonumber \\ \rho_{B,i} & = & p\rho_{i} + (1-p)  \delta_{i}, \label{ben}\eea which gives the same $p$ in all decompositions \cite{exp2} \bea p =\frac{1}{1+|\vec{n}| |\sin\theta | }. \label{p}\eea

We now apply Eq. (\ref{res}) to obtain a bound to the minimum-error state discrimination. Since each state $\rho_{j}$ is contained in the ensemble $\rho_{B,j}$ with the equal probability $p$ in (\ref{ben}) and (\ref{p}), the bound in (\ref{form}) can be applied,  \bea P_{s} \leq \frac{1}{N} (1+|\vec{n}| |\sin\theta| ). \label{sp} \eea This can also be expressed in terms of a lower bound to the minimum error, \bea P_e \geq 1- \frac{1+|\vec{n}| |\sin\theta |}{N}. \label{ep} \eea The bound above works for both pure ($|\vec{n}|=1$) and mixed ($|\vec{n}|<1$) states for any number $N$. Note that the optimization over ensemble decompositions, defined in (\ref{res}), is not performed here and a particular ensemble decomposition constructed in (\ref{ben}) is considered. In the next subsection, we show that the bound in Eq. (\ref{ep}) is in fact tight, meaning that the decomposition is also optimal for states in (\ref{nqstate}).

\subsection{The minimum-error discrimination}

We briefly review the optimal minimum-error discrimination among symmetric states. In a $D$ dimensional Hilbert space, $N$ symmetric states can be parameterized as follows, \bea |\phi_{j} \rangle = \sum_{k=1}^{D} c_{k} e^{\frac{2\pi i}{N} j(k-1)} |k\rangle, \label{sys}\eea with orthonormal basis $\{ |k\rangle \}$. Note that each state $|\phi_{j} \rangle$ is normalized, i.e. $\sum_{k} |c_{k}|^{2} = 1$. The optimal measurement that minimizes the error probability is known \cite{herzog}, \bea M_{j} = |\mu_{j} \rangle \langle \mu_{j}|, ~~~ |\mu_{j} \rangle = \frac{1}{\sqrt{N}} \sum_{k=1}^{D} \frac{c_{k}}{|c_{k}|} e^{\frac{2\pi i}{N} j (k-1)} |k\rangle. \label{ms} \eea It is therefore straightforward to compute the success probability, \bea P_{s} & = & \frac{1}{N}\sum_{i}P(~i~ |~ |\phi_{i}\rangle \langle \phi_{i}|) =  \frac{1}{N} \big(\sum_{k}|c_{k}| \big)^{2} . \label{mine}\eea For $D=2$ (i.e. qubits), $N$ symmetric (pure) states can be parameterized as, $ |\psi_{j} \rangle = \cos (\theta/2) |0 \rangle+ e^{\frac{2\pi i}{N}j} \sin (\theta/2) |1\rangle$. The success probability in discriminating among those $N$ states is then, \bea P_{s} = \frac{1}{N} (1+ |\sin\theta|), \eea which coincides with Eq. (\ref{sp}) for pure states (i.e., $|\vec{n}| = 1$). This also means that the decomposition taken in (\ref{ben}) is optimal such that the tight bound can be obtained.

The optimal discrimination among symmetric mixed states was analyzed by Chou and Hsu in Ref. \cite{ms}, where the optimal POVMs are constructed by solving the following equations, \bea \pi_{k} = V^{k} \Phi_{2}
|\varphi_{0}\rangle \langle \varphi_{0} | \Phi_{2}^{\dagger} (V^{\dagger})^{k},~~~~\Phi_{2} = \sum_{\lambda}c_{\lambda}|\lambda\rangle \langle \lambda| \nonumber \eea where $|\lambda\rangle$ are eigenvectors of $\Phi = \sum_{k} V^{k} |\varphi_{0}\rangle \langle \varphi_{0} | (V^{\dagger})^{k}$, $|\varphi_{0}\rangle$ is chosen such that $\langle\lambda|\varphi_{0}\rangle$ is real and non-zero for all $|\lambda\rangle$, and $c_{\lambda} = \langle \lambda | \varphi_{0}\rangle^{-1}$. Applying the above to the $N$ symmetric qubit states, it is straightforward to compute the optimal POVMs, which turn out to be the same as the case of pure states. It turns out that the minimum-error in state discrimination coincides with the success probability in Eq. (\ref{sp}). This also shows that the decomposition taken in (\ref{ben}) is optimal such that the tight bound can be obtained. \\

\textbf{\emph{ Example.}} We revisit the example shown in Ref. \cite{ms} as an application of the upper bound in Eq. (\ref{res}), the minimum-error discrimination among three symmetric mixed states: \bea \rho_{0} & = & \frac{1}{2} (I + \frac{1}{3} (-\sigma_{z})),\nonumber \\
\rho_{1} & = & V \rho_{0} V^{\dagger},~ ~\rho_{2} = V^{2} \rho_{0} (V^{\dagger})^{2}  \label{est} \eea with the $V$ operator that rotates a state by $2\pi/3$ about the $y$ axis in the Bloch sphere, \bea V = \left(\begin{array}{ccc}
\cos\frac{\pi}{3} & -\sin\frac{\pi}{3} \\
\sin\frac{\pi}{3} & \cos\frac{\pi}{3} \\
\end{array} \right). \nonumber \eea In Ref.\cite{ms}, the minimum-error was shown to be $5/9$ after optimization of the measurement basis. Now we explicitly derive the same value of minimum error, by plugging the following parameters, the purity of the state in Eq. (\ref{est}), $|\vec{n}| = 1/3$, the angle given $\theta=\pi/2$, and $N=3$, into Eq. (\ref{ep}): \bea P_{e} = 1 - \frac{1 + \frac{1}{3} \cdot 1}{3} = \frac{5}{9} \approx .5556. \nonumber \eea Hence, it is shown that the minimum-error can be obtained by the no-signaling principle.

One can also consider other known bounds for the states in the example above. In Ref. \cite{qiu}, known bounds are extensively compared with one another, and it turns out that the bound to the minimum-error, $L_{4}$ in \cite{qiu}, seems to be a relatively good one, \bea L_{4} = 1 - \min_{k}(\mu_{k} + \sum_{j\neq k} \tr(\mu_{j}\rho_{j} - \mu_{k}\rho_{k})_{+}), \label{l4}\eea where $\mu_{j}$ are \emph{a priori} probabilities, and $(A)_{+}$ denotes the positive part in the spectral decomposition of an operator $A$. One can consider the bound for the states in (\ref{est}) assuming that \emph{a priori} probabilities are equal. As it was shown in \cite{qiu}, the bound $L_{4}$ coincides with the optimal one if and only if, for some $k$ and all $j\neq l$, $(\rho_{j}-\rho_{k})_{+}$ and $(\rho_{l}-\rho_{k})_{+}$ are orthogonal. However, this condition is not fulfilled in the states in (\ref{est}). In fact, for those states, it holds that $\tr(\rho_{j} - \rho_{k})_{+} = 1/2\sqrt{3}$ for all $j\neq k$. Therefore, the bound $L_{4}$ is \bea L_{4} = 1 - (\frac{1}{3} +  \frac{1}{3}\frac{1}{2\sqrt{3}} +  \frac{1}{3}\frac{1}{2\sqrt{3}}) \approx .4742. \eea In addition, the bound proposed in Ref. \cite{qiu}, denoted by $L_{2}$ in Ref. \cite{qiu0}, is even worse as $L_{2} \approx .4519$.

\section{Examples in high dimensions}
\label{qudit}

It is remarkable that the result in Eq. (\ref{res}) does not assume particular dimensions of quantum states to be discriminated among. Therefore the bound in Eq. (\ref{res}) can be applied as long as an identical ensemble having different decompositions is constructed.
When applying the conditions (\ref{res}) and (\ref{form}) to derive a bound to the minimum-error state discrimination, one has to first construct the identical ensembles fulfilling the constraint in Eq. (\ref{ensem}). This is in fact non-trivial. For the cases of qubits, as it is shown in Sec. \ref{qubit}, the state space naturally has the $SO(3)$ picture, the Bloch sphere, based on which one can construct the identical ensembles. This, however, cannot be generalized to high dimensional systems due to the lack of representation as useful as the one in the two-dimensional case.

In what follows, we show an example of high dimensional quantum states to which the conditions in Eqs. (\ref{res}) and (\ref{form}) can be applied. The example generalizes the $SO(3)$ picture of the Bloch sphere to high dimensional systems.

\emph{$SO(3)$ states \cite{hp};} For quantum states of spin-$j$ particles, the basis are $|j, m\rangle$ where $m=j,-j+1,\cdots,j$. Generators for rotations are, $J_{1}$, $J_{2}$, and $J_{3}$ satisfying the commutation relation, \bea [J_{\alpha},J_{\beta}] = i \epsilon_{\alpha\beta \gamma} J_{\gamma}. \label{com} \eea Note that the generators can be written explicitly using the following elements, \bea
\langle j, (m+1) |~ (J_{1} + i J_{2}) |~ j, m \rangle & = & \sqrt{(j - m)(j + m + 1)} \nonumber \\
\langle j, (m+1) |~ (J_{1} - i J_{2}) |~ j, m \rangle & = & \sqrt{(j + m)(j - m + 1)} \nonumber \\
\langle j, m^{'} |~ J_{3} |~ j, m \rangle & = & m \delta_{m^{'},m}. \nonumber \eea For instance, for $j=1/2$ it holds that $J_{k} = \sigma_{k}/2$ where $k=1,2,3$ and $\sigma_{k}$ are Pauli matrices. A rotation operator can be represented with $J_{k}$ as $D(R) = D(n_{1},n_{2},n_{3}) = \exp(-i\theta \hat{n}\cdot \hat{J})$, that rotates a spin-$j$ state by $\theta$ with respect to the $\hat{J}$-axis.

Based on the rotation picture, identical ensembles of $\rho_{k}$ and $\delta_{k}$ can be constructed as follows. The states $\{ \rho_{k} \}$ for $k=0,\cdots,N-1$ that we wish to discriminate between are: \bea \rho_{0} & = & \frac{1}{2j+1} (I_{2j+1} + \vec{v}_{0} \cdot \hat{J}), ~~ \vec{v}_{0} = (-\alpha , 0 , \alpha), \nonumber \\
\rho_{k} & = & U_{k} \rho_{0} U_{k}^{\dagger}, ~~ U_{k} = \exp(-i\theta_{k} J_{3}),\label{rhok} \eea for some angles $\theta_{k}$ for each $\rho_{k}$. We also need the following states, \bea \sigma_{0} & = & \frac{1}{2j+1} (I_{2j+1} + \vec{w}_{0} \cdot \hat{J}), ~~ \vec{w}_{0} = (\beta , 0 , \beta), \label{si} \\
\sigma_{k} & = & U_{k} \sigma_{0} U_{k}^{\dagger}, ~~ U_{k} = \exp(-i\theta_{k} J_{3}),\label{sigma}\eea where the angles $\theta_{k}$ are given in Eq. (\ref{rhok}). In the above, $\alpha$ and $\beta$ are supposed to be given such that $\rho_{k}$ and $\sigma_{k}$ are non-negative. In the matrix representation, the operators $\rho_{k}$ and $\sigma_{k}$ are sparse matrices that can be easily diagonalized. The ensembles $\{ \rho_{B,k} \}$ identical for all $k$ are thus constructed, \bea \rho_{B,k} = p \rho_{k} + (1-p) \sigma_{k}, \nonumber \eea with $p=\beta / (\alpha+\beta)$, which points out to the $J_{3}$ direction. Note that the parameter $\beta$ from states $\sigma_{k}$ in (\ref{si}) is given only to construct the identical ensembles. Then, the formula in Eq. (\ref{form}) can be applied and immediately gives the bound,
\bea P_s \leq \min_{\beta} \frac{\alpha+\beta}{\beta N}. \label{fbd} \eea
It is thus shown by the no-signaling principle that the success probability in discrimination between states in Eq. (\ref{rhok}) is upper bounded.

\textbf{\emph{ Example.}} Let us consider a particular case of the states in (\ref{rhok}) that three states of spin-$1$ systems are given with equal probabilities. Then, let us take the following representation of generators, \bea J_{1} = \frac{1}{\sqrt{2}}\left(\begin{array}{ccc}
0 & 1 & 0 \\
1 & 0 & 1 \\
0 & 1 & 0 \\
\end{array} \right), ~~~~
J_{3} = \left(\begin{array}{ccc}
1 & 0 & 0 \\
0 & 0 & 0 \\
0 & 0 & -1 \\
\end{array} \right). \nonumber \eea
Then, $\rho_{1}$ and $\rho_{2}$ can be expressed by the rotation $U_{k} = diag[e^{-i\theta_{k}},1,e^{i\theta_{k}}]$, as $\rho_{k} = U_{k}\rho_{0}  U_{k}^{\dagger}$. In the same way, $\sigma_{k}$ in (\ref{si}) can be written, and $\sigma_{k}\geq 0$ for $\beta \in (0,1/\sqrt{2}]$. Therefore, the upper bound to the success probability from the no-signaling principle is, as it is shown in (\ref{fbd}), \bea P_{s} \leq \frac{1}{3} (1+ \sqrt{2}\alpha).\label{fbd1} \eea

Let us now compare the bound above with the success probability $1 - L_{4}$ in Ref.\cite{qiu}. As it was discussed, the bound is optimal if and only if for some $k$ and all $j\neq l$, $(\rho_{j}-\rho_{k})_{+}$ and $(\rho_{l}-\rho_{k})_{+}$ are orthogonal. However, this condition is not fulfilled in the states in (\ref{rhok}). In fact, for those states, it holds that $3 \tr(\rho_{j} - \rho_{k})_{+} = |\alpha|\sqrt{2 (1-\cos(\theta_{j}-\theta_{k}))} $ for all $j\neq k$. Since $\theta_{1} = 0$, we obtain \bea 1- L_{4} & = & \frac{1}{3}(1 + \eta(\theta_{1},\theta_{2})\alpha), \nonumber \\
\eta(\theta_{1},\theta_{2}) & = & \frac{2}{3}(\sin(\frac{\theta_{2}}{2}) + \sin(\frac{\theta_{3}}{2})). \label{l4s}\eea Note that $\eta(\theta_{1},\theta_{2}) < \sqrt{2}$ for all $\theta_{1}$ and $\theta_{2}$. Hence, it is shown that for three states in (\ref{rhok}) the bound $L_{4}$ is closer to the minimum-error than the one in (\ref{fbd1}). It is also clear that the bound in (\ref{fbd}) is not tight in general.

Finally, let us comment on why the bound in (\ref{fbd}) is not tight in contrast to the case of qubit states. For symmetric qubit states we have shown in Sec. \ref{qubit} that the tight bound can be obtained from the no-signaling condition. In fact, for the bound in (\ref{sp}) to be tight, it is necessary that the states in $(\ref{delta})$ to construct the identical ensemble in (\ref{ben}) are pure. Note that this was also mentioned in Ref. \cite{gen} in the context of generalized probability theories. Now, for the $SO(3)$ states, however, it is not fulfilled that the states in (\ref{si}) are pure, since $\tr[\sigma_{0}^{2}] = (3+4\beta^{2})/9 < 1$ for all $\beta \in (0,1/\sqrt{2}]$. Nevertheless, it cannot be excluded that a bound from the no-signaling principle is tight for $SO(3)$ or high-dimensional states unless one may consider the problem of finding the optimal ensemble decomposition as it is defined in (\ref{res}).

\section{Conclusion}
\label{conc}

We have derived an upper (lower) bound to the success (error) probability in the state discrimination by the no-signaling principle. The upper bound can be compared with known bounds to the success probability, e.g. Refs. \cite{monta,qiu,tyson1}, that are obtained by applying some inequalities to the minimum-error formula. The bound we provided here depends on neither dimensions nor particular structures of given quantum states. What is required to apply the bound in Eq. (\ref{res}) is to construct an identical ensemble that can have different ensemble decompositions such that each decomposition consists of each of quantum states to be discriminated among, as it is shown in Eq. (\ref{ensem}). As long as the ensemble satisfies the constraint, one can apply the bound in Eq. (\ref{res}) to a set of quantum states. It was also shown that the bound coincides with that of the minimum-error state discrimination for known cases such as $N$ symmetric (pure and mixed) qubit states. We derived a bound to minimum-error multiple state discrimination in high dimensions for which the optimal state discrimination is not known yet. Our results strengthen the quantitative connections among the no-go theorems, the no-cloning, the no perfect state estimation, and the no-signaling principle.

\section*{Acknowledgement}
We thank A. Ac\'\i n and T. Cubitt for useful discussions and R. Renner for discussions and comments on a connection to Ref. \cite{renner}. We also thank J. Tyson for useful comments and references. This study was financially supported by Chonnam National University 2009, the Korea Research Foundation Grant funded by the Korean Government under the contract number, KRF-2008-313-C00185, and the IT R$\&$D program of MKE/IITA (2008-F-035-01).


\begin{thebibliography}{99}
\bibitem{note} In the case that the probability is obtained by measuring quantum states, the probability $P(k|\rho_{l})$ can be expressed with POVMs, $\{ m_{k}\}$ that discriminates among $\{\rho_{k}\}$, $\sum_{k} m_{k} = I$, where each $m_{k}$ concludes that the state $\rho_{k}$ is detected: $P(k|\rho_{l}) = \tr[m_{k}\rho_{l}]$.
\bibitem{exp2} The value $p$ can be computed by taking the two identical ensembles $\rho_{B,i}$ and $\rho_{B,j}$ in (\ref{ben}) with $i\neq j$. It follows from the fact that  $\rho_{B,i} = \rho_{B,j}$, that $(1-p)(\delta_{j} - \delta_{i}) = p (\rho_{i} - \rho_{j})$. One can then apply the trace norm, denoted by $\| \cdot\|$, and obtain the following relation \bea p = \frac{\| \delta_{i} - \delta_{j} \|}{ \| \delta_{i} - \delta_{j} \| + \| \rho_{i} - \rho_{j} \|}. \label{pp}\eea Note that the trace norm is defined as, $\| A \| = \tr \sqrt{A^{\dagger}A}$, or equivalently, $\| A \| = \sum_{j}|\lambda_{j}|$ where $\lambda_{j}$ are eigenvalues of $A$. It is then explicitly computed in (\ref{p}).
\end{thebibliography}
\end{document}